\documentclass[12pt,epsf]{article}
\usepackage{amsmath, amssymb}
\usepackage{graphicx}
\begin{document}

\def\a{\alpha}
\def\b{\beta}
\def\c{\varepsilon}
\def\d{\delta}
\def\e{\epsilon}
\def\f{\phi}
\def\g{\gamma}
\def\h{\theta}
\def\k{\kappa}
\def\l{\lambda}
\def\m{\mu}
\def\n{\nu}
\def\p{\psi}
\def\q{\partial}
\def\r{\rho}
\def\s{\sigma}
\def\t{\tau}
\def\u{\upsilon}
\def\v{\varphi}
\def\w{\omega}
\def\x{\xi}
\def\y{\eta}
\def\z{\zeta}
\def\D{\Delta}
\def\G{\Gamma}
\def\H{\Theta}
\def\L{\Lambda}
\def\F{\Phi}
\def\P{\Psi}
\def\S{\Sigma}

\def\o{\over}
\def\beq{\begin{eqnarray}}
\def\eeq{\end{eqnarray}}
\newcommand{\gsim}{ \mathop{}_{\textstyle \sim}^{\textstyle >} }
\newcommand{\lsim}{ \mathop{}_{\textstyle \sim}^{\textstyle <} }
\newcommand{\vev}[1]{ \left\langle {#1} \right\rangle }
\newcommand{\bra}[1]{ \langle {#1} | }
\newcommand{\ket}[1]{ | {#1} \rangle }
\newcommand{\EV}{ {\rm eV} }
\newcommand{\KEV}{ {\rm keV} }
\newcommand{\MEV}{ {\rm MeV} }
\newcommand{\GEV}{ {\rm GeV} }
\newcommand{\TEV}{ {\rm TeV} }
\def\diag{\mathop{\rm diag}\nolimits}
\def\Spin{\mathop{\rm Spin}}
\def\SO{\mathop{\rm SO}}
\def\O{\mathop{\rm O}}
\def\SU{\mathop{\rm SU}}
\def\U{\mathop{\rm U}}
\def\Sp{\mathop{\rm Sp}}
\def\SL{\mathop{\rm SL}}
\def\tr{\mathop{\rm tr}}

\def\mi{m_{\phi}}
\def\mc{m_{\sigma}}
\def\mpl{M_{\rm pl}}
\def\en{_{{\rm end}}}
\def\os{_{{\rm osc}}}
\def\de{_{{\rm dec}}}
\def\rh{_{{ R}}}
\def\sw{_2}
\def\kpiv{k_{*}}
\def\piv{_{*}}

\def\IJMP{Int.~J.~Mod.~Phys. }
\def\MPL{Mod.~Phys.~Lett. }
\def\NP{Nucl.~Phys. }
\def\PL{Phys.~Lett. }
\def\PR{Phys.~Rev. }
\def\PRL{Phys.~Rev.~Lett. }
\def\PTP{Prog.~Theor.~Phys. }
\def\ZP{Z.~Phys. }


\baselineskip 0.7cm
\begin{titlepage}
\begin{flushright}
IPMU12-0201\\
ICRR-report-629-2012-18\\
\end{flushright}

\vskip 1.35cm
\begin{center}
{\large \bf 
Non-Gaussianity from Attractor Curvaton\\
}
\vskip 1.2cm
Keisuke Harigaya$^1$, Masahiro Ibe$^{2,1}$, Masahiro Kawasaki$^{2,1}$ and Tsutomu T. Yanagida$^1$
\vskip 0.4cm
$^1${\it Kavli IPMU, TODIAS, University of Tokyo, Kashiwa 277-8583, Japan}\\
$^2${\it ICRR, University of Tokyo, Kashiwa 277-8582, Japan}
\vskip 1.5cm

\abstract{
We propose a curvaton model in which the initial condition 
of the curvaton oscillation is determined 
by its attractor behavior during inflation.
Assuming a chaotic inflation model, 
we find that the initial condition determined by the attractor behavior
is appropriate to generate a sizable non-Gaussianity contribution to
the curvature perturbation, which will be  tested in the foreseeable future. 
Implications on the thermal  history of the universe and on 
particle physics models are also discussed.
}
\end{center}
\end{titlepage}

\setcounter{page}{2}

\section{Introduction}
\label{sec: introduction}
The origin of the large scale structures of the universe and the
fluctuation of the cosmic microwave background (CMB)
radiation can be successfully explained by 
the primordial density perturbation generated by cosmic 
inflation\,\cite{Guth:1980zm}\cite{Mukhanov:1981xt}. 
In particular, the simplest scenarios in which 
the density perturbation is dominated by  the fluctuation 
of a single inflaton field has been very successful so far.

In general, however, any light fields other than the inflaton also fluctuate
during inflation and  may contribute to the primordial density perturbation.
In fact, in the  curvaton mechanism\,\cite{Linde:1996gt}, 
the primordial density perturbation is  mainly generated not by the inflaton fluctuation
but by the fluctuation of a light field, the curvaton.
The curvaton model is attractive since 
the inflaton field does not need to provide the whole density perturbation any more,
which relaxes the constraints on the inflation models.

One of the interesting observable feature of the 
curvaton scenario is that it can lead to 
 large non-Gaussianity  in the density perturbation
when the curvaton energy density is subdominant at its decay time.
This is in contrast to the primordial density perturbation
generated by the inflaton fluctuation,
where the non-Gausssianity is predicted to be highly suppressed\,\cite{Maldacena:2002vr}.
Therefore, if the  large non-Gaussianity suggested by WMAP data\,\cite{Yadav:2007yy,Komatsu:2010fb}
is confirmed by the forthcoming Planck experiment \cite{Planck:2006aa},
the curvaton scenario becomes a more plausible explanation
for the density perturbation.

There is, however,  a drawback in the curvaton scenario if it 
generates sizable non-Gaussianity.
That is the initial condition problem of the oscillation of the curvaton field. 
As we will briefly review in the next section,
the initial field value of the curvaton oscillation is required to be about
a hundred times of the Hubble parameter during inflation
to generate sizable non-Gaussianity.
This initial field value of the curvaton is, however, difficult to be justified
since there seems no special meaning on such an intermediate scale.%
\footnote{
If the required initial field value were at the origin
of the curvaton field or at around the Planck scale,
for example, it could be rationalized.
That is, the origin of the curvaton can be a symmetry enhancement point,
which can be chosen during inflation if the symmetry
is also respected by the inflaton dynamics.
The field value at the Planck scale can be also chosen if the field value 
during inflation is determined stochastically\,\cite{Dimopoulos:2003ss},
since the scalar potential of the curvaton is expected to become very steep 
for the field value larger than the Planck scale.
}

In this paper, we propose a curvaton model in which the initial
condition of the curvaton oscillation is set by the so-called attractor
behavior (See Ref.~\cite{Linde:1984ti}, for example)
during inflation. 
As we will show, the initial condition fixed by the 
attractor behavior can be appropriate for the curvaton scenario
to account for the non-Gaussianity suggested by the WMAP  data.
We also discuss how high the decay temperature of the curvaton field can be.
The implications on the particle physics models are also discussed briefly.

This paper is organized as follows. In section\,\ref{sec: basics}, 
we briefly review the curvaton scenario. 
In section\,\ref{sec: model}, we discuss the curvaton model which has the attractor behavior.
There, we also discuss some implications of the model on the particle physics models.

\section{Brief Review of The Curvaton Scenario}
\label{sec: basics}
Before going to discuss the attractor behavior of the curvaton field, 
let us briefly review the basics of the curvaton scenario.
For simplicity, we assume a simple quadratic potential for the curvaton field, 
$\sigma$,
\begin{eqnarray}
 V(\sigma)=\frac{1}{2}\mc^2 \sigma^2\ ,
\end{eqnarray}
where $m$ denotes the mass of the curvaton.
In the curvaton scenario, the curvaton mass is assumed to be smaller than the Hubble parameter 
during inflation. 
We also assume that the Hubble induced mass
is suppressed.
With such a flat potential, the curvaton field is over-dumped and its field value is frozen to 
its initial value, $\sigma_i$, during inflation.  
It should be noted that the field perturbation around $\s_i$,  $\s = \s_i+\delta \s$, 
becomes the origin 
of the primordial density perturbation of the universe eventually. 

After inflation, 
the curvaton starts to oscillate around its vacuum value ($\s = 0$)
once the Hubble parameter falls below the mass of the curvaton,
i.e. $H \lesssim \mc$.
Eventually, the coherent oscillation of the curvaton decays into radiation 
at $t=t\de$.
Hereafter, we assume that the reheating process after inflation
completes before the decay of the coherent oscillation of the curvaton. 

In the curvaton scenario, the curvature perturbation $\zeta$ on spatial
slices of uniform density is evolving until the decay of the  curvaton
(see e.g. Ref.\,\cite{Lyth:2009zz} for a review).
After the decay of  the curvaton oscillation,
the curvature perturbation stops evolving and becomes constant 
at the super-horizon scale, which is given by \cite{Lyth:2002my},
\begin{eqnarray}
 \zeta &=&
 \left.  \frac{\rho_\sigma}{4\rho_r+3\rho_\sigma}
 \frac{\delta\rho_\sigma}{\rho_\sigma}
 \right|_{t = t\de}
  =\
   \left.  \frac{\rho_\sigma}{4\rho_r+3\rho_\sigma}\right|_{t=t\de}
 \left.
\frac{\delta \rho_\sigma}{\rho_\sigma}
\right|_{\rm{horizon\,\,exit}} \nonumber \\
&=&\frac{r\de}{4+3r\de}
  \left.
  \left(
\frac{2\delta \sigma}{\sigma_i}
+\frac{\delta \sigma^2}{\sigma_i^2}
\right)
\right|_{\rm{horizon\,\,exit}},
\label{eq:zeta0}
\end{eqnarray}
where $\r_{r}$ denotes the energy density of the radiation,
$\rho_{\s}$ and $\d\rho_{\s}$ the energy density of the curvaton and its fluctuation 
on the spatially flat slice.
In the above expression, $r\de$ is the ratio of the energy densities at the decay time of the curvaton,
$r\de=\rho_\sigma(t\de)/\rho_r(t\de)$, and we have used the fact that $\delta \rho_\s/\rho_\s$
is constant in time from the horizon exit to the decay time. 
It should be noted that  the inflaton contribution to the curvature perturbation is assumed to be
negligible.

By using  Eq.\,(\ref{eq:zeta0}), 
we obtain the power spectrum of
the curvature perturbation,
\begin{eqnarray}
 {\cal P}_\zeta(k)=\frac{4r\de^2}{(4+3r\de)^2}
 \left(\frac{H_k}{2\pi  \sigma_i}\right)^2
 \simeq
 \frac{r\de^2}{16\pi^2}\left(\frac{H_k}{\sigma_i}\right)^2\ ,
\label{eq:Pzeta0}
\end{eqnarray}
for $r\de \ll 1$.
Here, $H_k$ is the Hubble parameter at the horizon exit of the wave number $k$ 
and we have used $\delta \sigma_k = H_k/2\pi$. 
The nonlinearity parameter $f_{NL}$ is, on the other hand, 
defined by the bispectrum,
\begin{eqnarray}
 {\cal B}_\zeta({\bf k_1,k_2,k_3})=\frac{6}{5}f_{NL}(k_1,k_2,k_3)[{\cal
  P}_\zeta(k_1){\cal P}_\zeta(k_2)+~{\rm cyclic~permutations}]\ ,
\end{eqnarray}
which can be extracted from Eq.\,(\ref{eq:zeta0}) by remembering 
the relation to the Gaussian perturbation $\z_g$,
\begin{eqnarray}
\label{eq:fNL}
\zeta = \zeta_g +\frac{3}{5}f_{NL} 
 \zeta_g^2\ .
\end{eqnarray}
By comparing Eqs.\,(\ref{eq:zeta0}) and (\ref{eq:fNL}),
we find that 
\begin{eqnarray}
\zeta_g \simeq \frac{r\de} {4\pi} \frac{H_k}{\sigma_i} \ , \quad
f_{\rm NL} \simeq \frac{5}{3r\de}\ ,
\label{eq:fNL0}
\end{eqnarray}
for $r\de \ll 1$.
The above expression shows that  the nonlinearity parameter $f_{NL}$ can 
be sizable if the curvaton energy density at its decay time  is subdominant.

Now, let us estimate the ratio of the energy densities, $r\de$. 
Here, we assume that the curvaton starts to oscillate
before the reheating.
At the beginning of the oscillation\,\cite{Kawasaki:2011pd} set by,
\begin{eqnarray}
 H^2\simeq\left|\frac{1}{5\s}\frac{\partial V}{\partial \sigma}\right|=\frac{1}{5}\mc^2\equiv H\os^2\ ,
\label{eq:begin oscillation}
\end{eqnarray}
the ratio of the energy densities of the curvaton and the inflaton, $\phi$, is given by,
\begin{eqnarray}
\frac{\rho_{\s}}{\rho_{\phi}} \simeq \frac{V(\s_i)}{3M_{\rm pl}^2 H_{\rm osc}^2} 
\simeq \frac{5}{6}\frac{\s_i^2}{M_{\rm pl}^2}\ ,
\end{eqnarray}
which is a constant of time until the inflaton decays and the universe is reheated. 
Here, $M_{\rm pl}$ denotes the reduced Planck scale.
Once the reheating process completes, the energy density of the inflaton
is converted to the radiation energy density which is diluted faster than the 
curvaton density.
As a result, the ratio of the energy densities at the curvaton decay time is given by
\begin{eqnarray}
 r\de\simeq   
  \frac{5}{6}\frac{\sigma_i^2}{M_{\rm pl}^2}\times \frac{T_R}{T\de}\ , \,\,\, ({\rm for}\,\,\, T_R > T\de )\ ,
\label{eq:rdec0}
\end{eqnarray}
where $T_R$ and $T\de$ are the reheating temperature and the decay temperature
of the curvaton, respectively.

As a result, we find that  the initial value of the curvaton oscillation 
should be much smaller than the Planck scale, i.e. $\s_i \ll M_{\rm pl}$,
for a sizable nonlinearity parameter to be generated. 
More explicitly, 
we may express the required initial value $\sigma_i$ in terms of the power spectrum
and the nonlinearity parameter by using Eqs.\,(\ref{eq:Pzeta0}) and (\ref{eq:fNL0}),
\begin{eqnarray}
 \sigma_i&\simeq&\frac{5}{12 \pi}\frac{H_k}{{\cal
  P}_\zeta^{1/2}(k)f_{NL}}\nonumber\ ,\\
&\simeq& 70\times H_{k_{*}} \left(\frac{40}{f_{NL}}\right)
\left(\frac{4.9\times 10^{-5}}{{\cal
 P}_\zeta^{1/2}(k_*)}\right)\ ,
\label{eq:initial condition}
\end{eqnarray}
where $k_{*}=0.002{\rm Mpc}^{-1}$ is the pivot scale.
Here, we have again assumed $r\de\ll 1$.
As mentioned  in the introduction, 
this initial value of the curvaton is difficult to be rationalized,
since it seems no apparent physical meaning on that scale.
As we will discuss in the next section, however, 
the above initial condition is dynamically realized in the 
model with the attractor behavior.

Before closing this section, let us comment on the decay temperature, $T\de$.
First, let us remember that the baryon asymmetry of the universe and dark matter 
are required to be generated after  the curvaton decay
to avoid unacceptably large iso-curvature perturbations\,\cite{Lyth:2002my}.
Thus, if the baryon asymmetry is explained by thermal leptogenesis\,\cite{Fukugita:1986hr}, 
for example, 
the decay temperature of the curvaton is required to be high, $T\de\gtrsim 10^{9}$\,GeV,
since thermal leptogenesis requires 
a high temperature environment $O(10^{9})\,\GEV$\,\cite{Buchmuller:2004nz}.
On the other hand, from Eqs.\,(\ref{eq:fNL0}), (\ref{eq:rdec0}) and (\ref{eq:initial condition}),
the decay temperature is required to be
\begin{eqnarray}
 T\de&\simeq&\frac{25}{288\pi^2}\frac{1}{{\cal P}_\zeta(k)
  f_{NL}}\left(\frac{H_k}{\mpl}\right)^2T_R\ ,\nonumber\\
&\simeq&4 \times 10^{9} \GEV \left(\frac{H_{k_*}}{5\times 10^{13}{\rm GeV}}\right)^2
\left(\frac{40}{f_{NL}}\right)
\left(\frac{4.9\times 10^{-5}}{{\cal P}^{1/2}_\zeta(k_*)}\right)^2
\left(\frac{T_R}{10^{13}\GEV}\right)\  .
\end{eqnarray}
Thus, a rather high decay temperature, $T_{\rm dec}\gtrsim 10^9$\,GeV,
for example, can be achieved if the Hubble scale during inflation and the reheating 
temperature are rather high  for $f_{NL}= O(10)$.

Unfortunately, however, these requirements are not easily satisfied
in explicit inflation models.
To make our discussion concrete, 
let us consider the chaotic inflation model \cite{Linde:1983gd} with a quadratic potential 
$\frac{1}{2}\mi^2\phi^2$ as an example. 
In this model, the Hubble parameter is given by $H_k=\sqrt{2N_k/3}\mi$ where $N_k$ is the number
of $e$-foldings, and hence, the above $T\de$ is reduced to
\begin{eqnarray}
 T\de&\simeq&\frac{25}{432\pi^2}\frac{N_k}{{\cal P}_\zeta(k)
  f_{NL}}\left(\frac{\mi}{\mpl}\right)^2T_R\nonumber\ ,\\
&\simeq&0.7 \times 10^{6} \GEV \left(\frac{\mi}{10^{12} \GEV}\right)^3
\left(\frac{N_{k_*}}{60}\right)
\left(\frac{40}{f_{NL}}\right)
\left(\frac{4.9\times 10^{-5}}{{\cal P}^{1/2}_\zeta(k_*)}\right)^2
\left(\frac{T_R}{\mi}\right)\ .
\end{eqnarray}
It should be noted that the inflaton mass cannot be much larger than $m_\phi = 10^{12}$\,GeV,
since otherwise the inflaton contribution to the 
curvature perturbation, 
\begin{eqnarray}
{\cal P}_\zeta(k_*)^{\rm inflaton}\simeq \frac{N_{k_*}^2 m_\phi^2}{6\pi^2 M_{\rm pl}^2}\ ,
\end{eqnarray}
cannot be ignored and  the non-Gaussianity is suppressed.%
\footnote{
If the inflaton contribution to the curvature perturbation is sizable, 
baryogenesis before the curvaton decay 
can be consistent with the current constraint on the isocurvature fluctuation\,\cite{Langlois:2008vk}. 
}
Therefore, the decay temperature is expected to be rather low 
for a sizable nonlinearity  parameter, $f_{NL} = O(10)$,
unless the reheating temperature after inflation is much 
higher than the inflaton mass (see Refs.~\cite{Kolb:2003ke}
for discussions on the reheating temperature and the inflaton mass).

\section{Curvaton Model with Attractor Behaviors}
\label{sec: model}
Now, let us discuss the curvaton model in which the 
curvaton field value after inflation is set by the attractor behavior.
To be concrete,
we consider the chaotic inflation model with a curvaton  field
whose scalar potential is given by,
\begin{eqnarray}
 V(\phi,\sigma)=\frac{1}{2}\mi^2\phi^2+\frac{1}{4}\lambda \sigma^4
  + \frac{1}{2} \mc^2 \sigma^2\ .
\end{eqnarray}
where $\l$ denotes a coupling constant.
We assume that the Hubble induced mass is negligible.
Notice that the spectral index $n_s$ of the curvature perturbation is
red-tilted ($n_s\simeq 0.98$) in the chaotic inflation scenario, which is
consistent with the CMB observations \cite{Komatsu:2010fb}.
\subsection{Dynamics of the fields during  inflation}
When the slow-roll conditions are satisfied, the dynamics of the
inflaton field and the curvaton field are described by 
\begin{eqnarray}
\label{eq:EOM}
 3H\dot{\phi}=-\partial_\phi V,~~ 3H\dot{\sigma}=-\partial_\sigma V\ .
\end{eqnarray}
In the followings, we assume that the initial value of the 
inflaton at the beginning of inflation, $\phi_0$,  is much larger than the Planck scale
since we are considering the chaotic inflation scenario.
The initial value of the curvaton, $\s_0$, is, on the other hand,  at around the Planck scale,
which can be justified if the curvaton potential steeply increases for $\s \gtrsim M_{\rm pl}$.

\begin{figure}[tb]
\begin{center}
  \includegraphics[width=.48\linewidth]{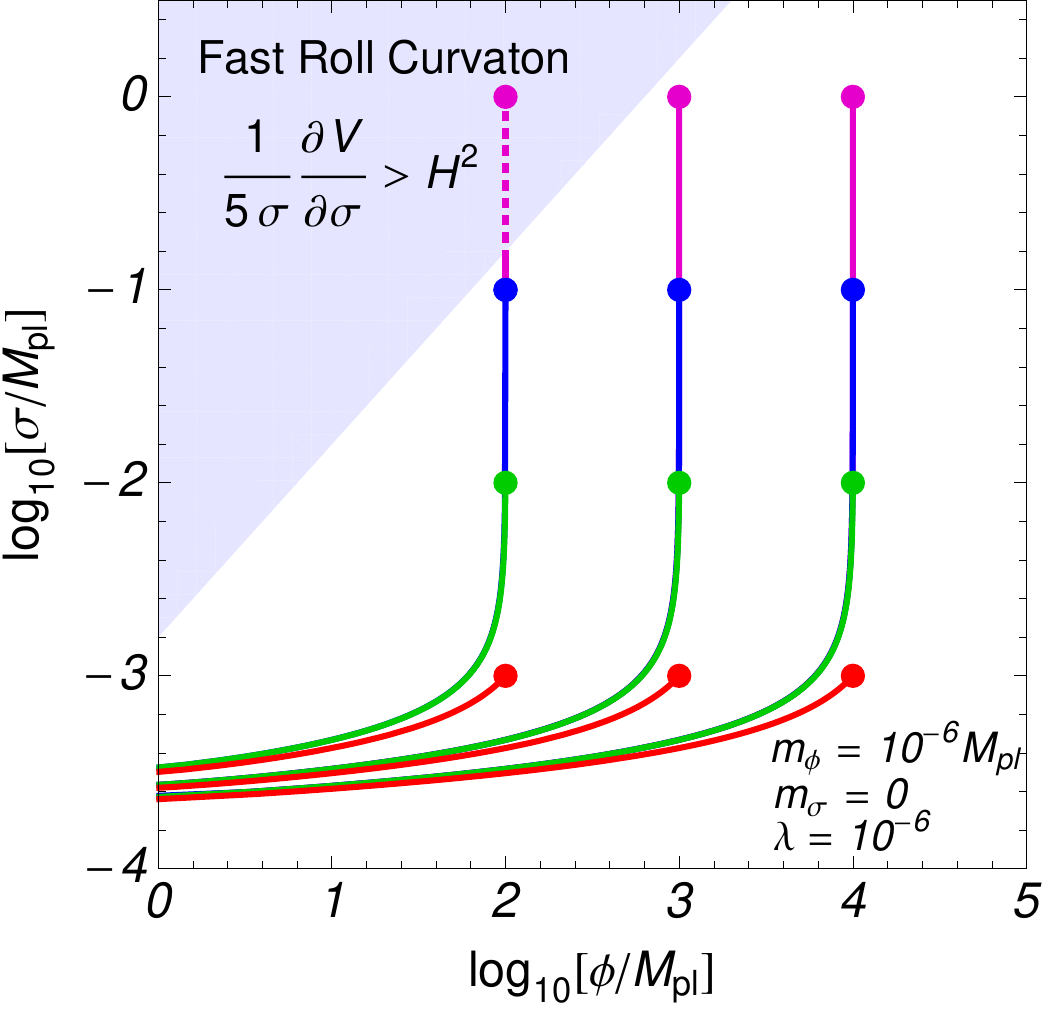}
 \end{center}
\caption{\sl \small
The attractor behavior of the curvaton field during inflation.
The magenta, blue, green and red lines correspond to the 
initial condition $\s_0 = 1,10^{-1},10^{-2},10^{-3}\times M_{\rm pl}$, respectively,
while the initial condition of $\phi_0$ is $10^2, 10^3, 10^4\times M_{\rm pl}$
from left to right.
The figure shows that the curvaton becomes insensitive to its initial condition.
In the colored region, slow-roll condition for the curvaton field is
 violated and the curvaton field promptly rolls down until it gets out of
 the colored region, as is shown with the dotted line. Afterwards, the curvaton field follows the attractor
 solutions shown in the figure.
}
\label{fig:attractor}
\end{figure}

In the large curvaton field value region, $\sigma>\sqrt{2}\mc/\sqrt{\lambda}\equiv \sigma\sw$, 
the potential of the curvaton field is dominated by the quartic term. 
In this region, the solutions of the equations of motions of the inflaton and the curvaton (Eq.\,(\ref{eq:EOM}))
are interrelated by,
\begin{eqnarray}
\frac{1}{\mi^2}{\log}\frac{\phi_0}{\phi}= \frac{1}{2\lambda}\left(\frac{1}{\sigma^2}-\frac{1}{\sigma_0^2}\right)\ ,
\end{eqnarray}
where the subscript ``$0$'' denotes the initial values 
at the beginning of inflation.
After enough time,
$\sigma$ becomes far smaller than $\sigma_0$,
and hence, the curvaton field
enters into the attractor solution given by
\begin{eqnarray}
 \sigma_{\rm att}(\phi)=\frac{\mi}{\sqrt{2\lambda}}
 \left({\log}\frac{\phi_0}{\phi}\right)^{-1/2}\ .
\label{eq:attractor}
\end{eqnarray}
In Fig.\,\ref{fig:attractor}, we show an attractive behavior of the
massless curvaton during inflation for a small initial value
$\sigma_0<\mpl$.
The figure shows that the curvaton evolves along the attractor solution and becomes
insensitive to its initial value $\sigma_0$. 

When the curvaton field becomes smaller than $\s < \s_2$ during inflation,
the evolution of the curvaton deviates from the attractor solution,
while it keeps following the attractor behavior by 
the end of inflation if the curvaton field satisfies $\s > \s_2$ until then.
In both cases, the initial condition of the curvaton oscillation after inflation, $\s_i$,
is insensitive to its initial condition at the beginning of inflation, $\s_0$.
As we will show shortly, 
the dynamically chosen $\s_i$ can be appropriate to generate 
a large $f_{NL}$, and hence, the model with the attractor behavior provides us 
with a solution to the initial condition problem.

Before proceeding to the details of the curvaton behaviors, 
we comment on the initial condition of the inflaton field. 
It  should be noted that the initial condition of the inflaton cannot be
arbitrarily large since the stochastic behavior is significant 
for  $\dot{\phi}H^{-1}<H/{2\pi}$. 
For the quadratic potential, this condition corresponds to
\begin{eqnarray}
\phi>(96\pi^2)^{1/4}
\mpl^{3/2}/\mi^{1/2}
\sim 10^4 \times \mpl 
\left(\frac{10^{12}
\, \GEV}{\mi}\right)^{\frac{1}{2}}. 
\label{eq:stochastic region}
\end{eqnarray}
If the universe begins when the inflaton escapes 
from the stochastic region, the initial condition is given by 
the field value which saturates the inequality of Eq.\,(\ref{eq:stochastic region}).
In the followings, we assume that the initial condition of the
inflaton field is given by the saturated value,
\begin{eqnarray}
\phi_0 \sim 10^4 \times \mpl 
\left(\frac{10^{12}
\, \GEV}{\mi}\right)^{\frac{1}{2}} \ ,
\end{eqnarray}
although the attractive behavior is less sensitive 
 to the initial condition of the inflaton, $\phi_0$ (see Fig.\,\ref{fig:attractor}).

\subsection{ $\s_2 < \s_{\rm att}(\phi_{\rm end})$}
\label{sec: case1}
When the curvaton mass is small enough, 
it follows the attractor solution
by the end of inflation,
\begin{eqnarray}
 \phi\simeq 2 \mpl \equiv \phi\en . 
\end{eqnarray}
Thus, in this case, the initial condition of the curvaton after inflation is given by,%
\footnote{The condition that the curvaton
keeps following the attractor solution by the end of inflation
is given by $m_\s < m_\phi/(2(\log(\phi_0/\phi\en))^{1/2})$.}
\begin{eqnarray}
\s_i\simeq \s\en \equiv \frac{\mi}{\sqrt{2\lambda}} 
 \left({\log}\frac{\phi_0}{\phi_{\rm end}}\right)^{-1/2}\ .
\end{eqnarray}
After the end of inflation, the inflaton field starts oscillation around the
minimum of the potential. 
By remembering that 
\begin{eqnarray}
 H\en^2&\simeq&\frac{1}{3\mpl^2}\frac{1}{2}\mi^2
  \phi\en^2=\frac{2}{3}\mi^2\ ,\nonumber\\
\left.\frac{1}{5\s} \frac{\partial V}{\partial \s}
\right|\en&\simeq&\frac{\lambda}{5}\sigma\en^2=\frac{1}{10}\mi^2 x\en^{-1}<H\en^2\ ,
\end{eqnarray}
at the end of inflation,
we see that the oscillation condition in Eq.\,(\ref{eq:begin oscillation})
is satisfied just a few Hubble times after the end of inflation.
Here, we have defined $x\en = \log(\phi_0/\phi\en)$.
Thus, it is reasonable to assume that the curvaton oscillation starts  before the
reheating of the universe.

When the curvaton field starts oscillation, the potential of the curvaton is governed by the quartic term. 
In this period, the energy density of the curvaton oscillation scales as the energy density of the radiation.
The power spectrum of the curvature perturbation in such a case is given in
Ref.~\cite{Kawasaki:2011pd} (see also the appendix~\ref{sec: formulas}), 
\begin{eqnarray}
 {\cal P}_{\zeta}(k)\simeq
\frac{9r\de^2}{(4+3r\de)^2}\frac{\sigma\os^4}{\sigma_k^6}
\left(\frac{H_k}{2\pi}\right)^2 \ .
\end{eqnarray}
Here, $\sigma\os$ and $\sigma_k$ are the field values of the curvaton 
at the beginning of the curvaton oscillation and at the 
horizon exit of the wave number $k$ during inflation, respectively.
As we have mentioned above, 
the curvaton starts oscillation just after the end of inflation,
and hence,
$\sigma\os \sim \sigma_i$, while $\s_k$ is given by
$\sigma_k=\mi/\sqrt{2\lambda} \log(\phi_0/\phi_k)^{-1/2}$ with 
$\phi_k$ being the field value of the inflaton 
at the horizon exit of the wave number $k$.
As a result, the power spectrum is given by
\begin{eqnarray}
 {\cal P}_{\zeta}(k)&=&
  \frac{3r\de^2}{(4+3r\de)^2}\frac{\lambda}{\pi^2}N_kx\en^{-2}x_k^3\nonumber\\
&\simeq& 2 \times 10^{-9}
 \left(\frac{r\de}{0.26}\right)^2
\left( \frac{\lambda}{10^{-8}}\right)
\left(\frac{N_k}{60}\right)
\left(\frac{9}{x\en}\right)^{2}
\left(\frac{x_k}{7}\right)^3 \ ,
\label{eq:Pzeta1pre}
\end{eqnarray}
where we have defined $x_k = \log(\phi_0/\phi_k)$
and assumed that $r\de\ll 1$.
The nonlinearity parameter $f_{NL}$ is also given by 
\begin{eqnarray}
f_{NL}&=&\frac{16(1+r\de)}{r\de(4+3r\de)}
+\frac{4+3r\de}{r\de}\left(2-\frac{\sigma_k^2}{\sigma\os^2}\right)\ ,\nonumber\\
&\simeq&\frac{4}{r\de}\left(3-\frac{x\en}{x_k}\right) \ ,
\label{eq:fnl1}
\end{eqnarray}
for $r\de\ll 1$ (see also the appendix\,\ref{sec: formulas}).

Finally, let us estimate the energy fraction of the curvaton at the decay time, $r\de$. 
The energy densities of the radiation
and the curvaton at the curvaton decay time are 
given by
\begin{eqnarray}
 \rho_r|\de&=&\rho_\phi|\en
 \times\left(\frac{a\en}{a\rh}\right)^3
 \left(\frac{a\rh}{a\de}\right)^4\ ,
  \nonumber\\
 \rho_\sigma|\de&=&\rho_\sigma|\en\times
 \left(\frac{a\os}{a\sw}\right)^4\left(\frac{a\sw}{a\de}\right)^3\ ,
\end{eqnarray}
where $a$'s are the scale factors of the universe.
The 
meanings of the subscripts are understood.
Therefore, $r\de$ can be represented in terms of the scale factors as
\begin{eqnarray}
 r\de=
\left(  \frac{a\de}{a\rh}\right)
\left(\frac{a\os}{a\sw}\right)
\left(\frac{a\os}{a\en}\right)^3 \times r\en\  ,
\end{eqnarray}
with 
\begin{eqnarray}
r\en = \left.\frac{\r_\s}{\r_r}\right|_{\rm end} \ .
\end{eqnarray}
The first factor is given by $a\de/a\rh=T\rh/T\de$, 
since the temperature of the universe is inversely proportional to the scale
factor after the reheating.
Between
$t\os$ and $t\sw$, the amplitude of the curvaton field 
is inversely proportional to the scale factor,
and hence, we obtain
$a\os/a\sw=\sigma\sw/\sigma_i
=2\mc/\mi\times\sqrt{x\en}$.
The third factor is, on the other hand, given by 
\begin{eqnarray}
 \left(\frac{a\os}{a\en}\right)^3
 =\frac{H\en^2}{H\os^2}
 =
 \left.H\en^2\left(
 \frac{1}{5\s}
 \frac{\partial V}{\partial\sigma}\right)^{-1}\right|\en
\simeq\frac{20}{3}x\en\ .
\end{eqnarray}
The curvaton energy fraction at the end of inflation, $r\en$, is simply given by
\begin{eqnarray}
 r\en = \frac{\frac{1}{4}\lambda
  \sigma\en^4}{\frac{1}{2}\mi^2\phi\en^2}=\frac{1}{32 \lambda}
  \left(\frac{\mi}{\mpl}\right)^2x\en^{-2}\ .
\end{eqnarray}
Putting all of them together, $r\de$ is given by
\begin{eqnarray}
 r\de&\simeq&
 \frac{5}{12\lambda} \frac{\mc\mi}{\mpl^2}\frac{T\rh}{T\de}x\en^{-1/2}\ ,
 \\
 \label{eq:rdecay}
&\simeq&
 0.26\times
 \left(\frac{\mc}{10^{11}\,{\rm GeV}}\right)
 \left(\frac{\mi}{10^{12}\,{\rm  GeV}}\right)
  \left(\frac{T\rh}{10^{12}\,{\rm GeV}}\right)
 \left( \frac{10^{6}\,{\rm  GeV}}{T\de}\right)
 \left(\frac{10^{-8}}{\lambda}\right)
  \left(\frac{8}{x\en}\right)^{1/2}\ \nonumber.
\end{eqnarray}

As a result, by substituting the above $r\de$ into Eq.\,(\ref{eq:Pzeta1pre}), 
we obtain,
\begin{eqnarray}
P_{\zeta}(k)&\simeq&
2\times 10^{-9} \left(\frac{\mc}{10^{11}{\rm GeV}}\right)^2 
\left(\frac{\mi}{10^{12}{\rm GeV}}\right)^2
\left(\frac{T\rh}{10^{12}{\rm GeV}}\right)^2
 \left(\frac{10^{6}{\rm GeV}}{T\de}\right)^2\nonumber\\
&&
\times
\left(\frac{10^{-8}}{\lambda}\right)
\left(\frac{x_k}{7}\right)^3
\left(\frac{9}{x\en}\right)^3
\left(\frac{N_k}{60}\right)\ ,
\label{eq:Pzeta1}
\end{eqnarray}
while the nonlinearity parameter in Eq.\,(\ref{eq:fnl1}) is of $O(10)$ 
for the above parameter region.
Therefore, we find that the curvaton model where the initial amplitude of its oscillation
is set
by the attractor behavior can explain the observed power spectrum
while predicting a rather large nonlinearity parameter.
We should again emphasize that the above result is insensitive to the initial 
condition of the curvaton field as long as it follows the attractor 
behavior during inflation.

\subsection{ $\s_{\rm att}(\phi_{\rm end})<\s_2$}
\label{sec: case2}
For a larger curvaton mass,  $m_\s > m_\phi/(2\sqrt x\en)$,
the curvaton field  reaches $\s_2$ during inflation.%
\footnote{For $m_\s > m_\phi$, the curvaton field is settle to its 
minimum at $\s = 0$ during inflation, and hence,  we assume $m_\s < m_\phi$ in our analysis.}
In this case, the curvaton stops following the attractor behavior during inflation
and stays at $\sigma\simeq\sigma\sw$ until it starts oscillation. 
When the Hubble parameter decreases below the curvaton mass (see Eq.\,(\ref{eq:begin oscillation})),
the curvaton field starts oscillation as a massive field.
The power spectrum for such a curvaton is 
given in the appendix\,\ref{sec: formulas} (for $n=3$, $V=\frac{1}{4}\lambda\sigma^4
+ \frac{1}{2}\mc^2\sigma^2$ and $\sigma\os=\sigma\sw$), 
\begin{eqnarray}
 {\cal P}_\zeta(k)
 &\simeq&\frac{r\de^2}{(4+3r\de)^2}\frac{9\lambda}{2\mc^2}\left(\frac{H_k}{2\pi}\right)^2
 \simeq\frac{3r\de^2}{4(4+3r\de)^2}\frac{\lambda}{\pi^2}\left(\frac{\mi}{\mc}\right)^2N_k,\nonumber\\
&\simeq&2.5\times 10^{-9}
\left(\frac{r\de}{0.1}\right)^2
\left(\frac{\lambda}{10^{-6}}\right)
\left(\frac{N_k}{60}\right)
\left(\frac{\mi}{\mc}\right)^2\ .
\label{eq:Pzeta2pre}
\end{eqnarray}
The nonlinearity parameter is also given by,
\begin{eqnarray}
  f_{NL}&\simeq&
  \frac{40(1+r\de)}{3r\de(4+3r\de)}+\frac{5(4+3r\de)}{6r\de}\frac{86}{225}\ ,\nonumber\\
&\simeq&\frac{622}{135\,r\de}\ .
\label{eq:fnl2}
\end{eqnarray}

By remembering that the energy densities at the curvaton decay time are given by
\begin{eqnarray}
 \rho_\phi|\de&\simeq&\rho_\phi|\en
 \left(\frac{a\en}{a\rh}\right)^3
 \left(\frac{a\rh}{a\de}\right)^4\ ,
  \nonumber\\
 \rho_\sigma|\de&\simeq&\rho_\sigma|\en\left(\frac{a\sw}{a\de}\right)^3 \ ,
\end{eqnarray}
we find that the energy ratio can be expressed in terms of the scale factor of the universe,
\begin{eqnarray}
r\de\simeq \left(\frac{a\os}{a\en}\right)^3\left(\frac{a\de}{a\rh}\right) \times r\en \ .
\end{eqnarray}
As a result, by using the ratios of the scale factors obtained in the previous section,
we find
\begin{eqnarray}
 r\de\simeq\frac{10}{9\lambda}\left(\frac{\mc}{\mpl}\right)^2\frac{T\rh}{T\de}
 \simeq 0.1 \times
 \left(\frac{10^{-6}}{\lambda} \right)
\left(\frac{m_\s}{10^{12}\GEV} \right)^2
\left(\frac{T_R}{10^{12}\GEV} \right)
\left(\frac{2\times10^{6}\GEV}{T\de} \right)
\ .
\label{eq:rdec2}
\end{eqnarray}
Therefore, by putting $r\de$ into the above power spectrum in Eq.\,(\ref{eq:Pzeta2pre}),
we obtain,
\begin{eqnarray}
P_{\zeta}(k)&\simeq&
2\times 10^{-9} \left(\frac{\mc}{10^{12}{\rm GeV}}\right)^2 
\left(\frac{\mi}{10^{12}{\rm GeV}}\right)^2
\left(\frac{T\rh}{10^{12}{\rm GeV}}\right)^2
 \left(\frac{2\times10^{6}{\rm GeV}}{T\de}\right)^2\nonumber\\
&&
\times
\left(\frac{10^{-6}}{\lambda}\right)
\left(\frac{N_k}{60}\right)\ ,
\label{eq:Pzeta2}
\end{eqnarray}
while obtaining a rather large nonlinearity parameter  $f_{NL} = O(10)$ (see Eq.\,(\ref{eq:fnl2})).

\subsection{Implications on particle physics models}
Finally, let us discuss some implications of the model.
First, let us comment on the decay temperature of the curvaton field.
As we have discussed in section\,\ref{sec: basics} based on 
the simplest curvaton model, 
the decay temperature cannot be very high 
if we require a rather large nonlinearity parameter, $f_{NL} = O(10)$.
Here, we show that the similar conclusions are reached 
in the present model.
To see such a constraint explicitly, let us express the decay temperature $T\de$ in terms
of $P_\zeta$ and $f_{NL}$ by using Eq.~(\ref{eq:fnl2}), (\ref{eq:rdec2})
and (\ref{eq:Pzeta2}),%
\footnote{
Eqs.\,(\ref{eq:Pzeta1}) and (\ref{eq:Pzeta2}) show that the 
higher decay temperature is allowed for the higher curvaton mass.
Since we are interested in the upper bound on the required decay temperature,
we concentrate on the region $\mi/(2\sqrt{x\en})<\mc<\mi$ in the following arguments.
}
\begin{eqnarray}
 T\de&\simeq&
  \frac{311N_k}{1296\pi^2}\frac{T\rh}{\mi}\frac{\mi^3}{\mpl^2}\frac{1}{P_{\zeta}(k)f_{NL}}\ ,
  \nonumber\\
&\simeq& 3\times 10^{6}\,{\rm GeV} 
\left(\frac{\mi}{10^{12}{\rm GeV}}\right)^3
\left(\frac{40}{f_{NL}}\right)
\left(\frac{T\rh}{\mi}\right)
\left(\frac{N_{\kpiv}}{60}\right)
\left(\frac{4.9\times 10^{-5}}{{\cal P}_\zeta(\kpiv)^{1/2}}\right)^2\ .
\end{eqnarray}
This shows that the required decay temperature 
is again rather suppressed for a sizable nonlinearity parameter,
although it is slightly larger than the one
given in section\,\ref{sec: basics} by about a factor of 4.

Secondly, let us discuss a possible interrelation of the present model to 
another well-motivated new physics model, the see-saw mechanism\,\cite{seesaw}.
For that purpose, 
let us remember that the required quartic coupling $\lambda$ 
can be expressed by
(see Eqs.\,(\ref{eq:Pzeta1}) and (\ref{eq:Pzeta2})),
\begin{eqnarray}
 \lambda&\simeq&\frac{97200\pi^2}{96721N_k} f_{NL}^2 {\cal
  P}_{\zeta}(k)
  \left(\frac{\mc}{\mi}\right)^2\ ,\nonumber\\
&\simeq&6\times 10^{-7}
\left(\frac{f_{NL}}{40}\right)^2
\left(\frac{\mc}{\mi}\right)^2
\left(\frac{60}{N_{\kpiv}}\right)
\left(\frac{{\cal P}^{1/2}_\zeta(k)}{4.9\times 10^{-5}}\right) \ .
\end{eqnarray}
If we assume that the mass of the curvaton field
is generated by the vacuum expectation value of a field $X$
in a supersymmetric model via a superpotential,
\begin{eqnarray}
 W=g X \Sigma \Sigma\ ,~~\sigma=\frac{1}{\sqrt{2}}{\rm Re}(\Sigma)\ ,
\end{eqnarray}
the mass and the quartic coupling of the curvaton
are given by $\mc = g\vev X$ and $\l = g^2$, respectively. 
Thus, for example, to realize $\lambda\sim 10^{-6}$ and $\mc\sim10^{12}$ GeV,
the vacuum expectation value of $X$ is $\vev{X}\sim 10^{15}$ GeV. 
It is quite suggestive that  this value is close to the mass scale of the seesaw mechanism.
Actually, it has been discussed that the right-handed scalar neutrinos can play  a role of the curvaton field
in the supersymmetric seesaw mechanism\,\cite{Moroi:2002vx}. 
Detailed studies of the attractor behavior right-handed scalar neutrinos
as well as the compatibility with leptogenesis will be given
elsewhere\,\cite{HIKY}.

Thirdly, let us comment on an implication on compensated
isocurvature perturbations \cite{Gordon:2002gv}. If the dark matter abundance is created before the
curvaton decays, unacceptably large matter isocurvature perturbations are
generated and contradicts with the recent observations of
the CMB \cite{Komatsu:2010fb}.
This can happen in the models of dark matter such as
the gravitino produced by thermal scatterings or its decay products \cite{WinoDM},
the Q-ball or its decay products \cite{Kusenko:1997si,Enqvist:1997si},
the WIMPZILLA \cite{Chung:1998zb} and
the primordial black hole \cite{Hawking:1971ei}.
However, the constraint can be relaxed if the baryon
asymmetry is created as the curvaton decays, since the dark matter and
the baryon isocurvature perturbations compensate for each other.
Such scenario is possible if the right-handed scalar neutrino is the
curvaton.
The compensation requires $r\de=\Omega_b/\Omega_{\rm DM}\simeq0.2$
\cite{Gordon:2002gv}, where $\Omega_b$ and $\Omega_{\rm DM}$ are the
energy density fraction of the baryon and the dark matter, respectively,
and hence the nonlinearity parameter is predicted in each curvaton models.
For a curvaton model with
a quadratic potential, from Eq.~(\ref{eq:fNL0}), $f_{\rm NL}\simeq 8$.
In the case of an attractor curvaton with a quartic
potential, from Eqs.~(\ref{eq:fnl1}) and (\ref{eq:fnl2}), $f_{\rm NL}$ can be
as large as 40.

As a final remark, let us comment on the Hubble induced mass. In this
paper, we have assumed that the Hubble induced mass is
negligible. 
In supergravity, scalar fields obtain soft masses as
large as the Hubble scale if inflation is driven by $F$-terms \cite{Ovrut:1983my},
unless there exists a tuning of few percents, a Heisenberg symmetry
\cite{Murayama:1993xu}, a discrete $R$ symmetry \cite{Kumekawa:1994gx} or a shift symmetry \cite{Kawasaki:2000yn}, or scalar fields are pseudo
Nambu-Goldstone bosons \cite{ArkaniHamed:2003mz}. It will be
interesting to construct curvaton models with an attractor behavior but
without fine-tuning in supergravity.


\section*{Acknowledgments}
This work is supported by Grant-in-Aid for Scientific research from the
Ministry of Education, Science, Sports, and Culture (MEXT), Japan, No. 14102004 (M.K.), No. 21111006 (M.K.), No.\ 22244021 (T.T.Y.),
No.\ 24740151 (M.I),  and also by World Premier International Research Center Initiative (WPI Initiative), MEXT, Japan.
 The work of K.H. is supported in part by a JSPS Research Fellowships for Young Scientists.

\appendix
\section{General formulas for the power spectrum and the  non-Gaussianity}
\label{sec: formulas}
In this appendix, we summarize the formulas for the curvature
perturbation and the non-Gaussianity given in
Ref.\,\cite{Kawasaki:2011pd}.
The assumptions are
\begin{list}{$\cdot$}{}
 \item
The energy density of the curvaton oscillation is negligible compared to the total
      energy density at least until the onset of the curvaton oscillation.
 \item
The curvaton field starts oscillation suddenly at $t=t\os$ and its energy density scales
by $a^{-n}$ in terms of the scale factor of the universe $a$.
 \item
At some point before the curvaton  decays, the quadratic term
 dominates and the energy density of the curvaton begins to scale by $a^{-3}$.
 \item
The curvaton oscillation decays suddenly and the curvature perturbation is fixed at
      that point.
 \item
The curvature perturbation  from the fluctuation of the inflaton field is negligible.
 \item
The energy density of the inflaton field scales by $a^{-3}$
      after the end of inflation until the reheating.
\item
The reheating completes instantaneously at $t=t\rh$ before the curvaton  decays.
\end{list}
With these assumptions, the power spectrum of the curvature perturbation
is given by
\begin{eqnarray}
\label{eq:powerspectrumA}
 {\cal P}_\zeta(k)^{\frac{1}{2}}=\frac{r\de}{4+3r\de}(1-X(\sigma\os))^{-1}\frac{V'(\sigma\os)}{V'(\sigma_k)}\left(\frac{3}{n}\frac{V'(\sigma\os)}{V(\sigma\os)}-\frac{3X(\sigma\os)}{\sigma\os}\right)
 \frac{H_k}{2\pi}\ ,
\end{eqnarray}
where the function $X$ is defined by 
\begin{eqnarray}
 X(\sigma)=\frac{1}{2(c-3)}
 \left(\frac{\sigma
  V''(\sigma)}{V'(\sigma)}-1\right)\ ,\nonumber\\
c=
\left\{
\begin{array}{ll}
 5& (t\rh<t\os) \ , \\
{9}/{2}& (t\os<t\rh) \ .\\
\end{array}
\right.
\end{eqnarray}
The nonlinearity parameter is given by
\begin{eqnarray}
 f_{NL}(k)&=&\frac{40(1+r\de)}{3r\de
  (4+3r\de)}+\frac{5(4+3r\de)}{6r\de}
  \left(\frac{3}{n}\frac{V'(\sigma\os)}{V(\sigma\os)}-\frac{3X(\sigma\os)}{\sigma\os}\right)^{-1}
  \times\nonumber\\
&&\left[\frac{X'(\sigma\os)}{1-X(\sigma\os)}+
\left(\frac{3}{n}\frac{V'(\sigma\os)}{V(\sigma\os)}-\frac{3X(\sigma\os)}{\sigma\os}\right)^{-1}
\left\{\frac{3}{n}\frac{V''(\sigma\os)}{V(\sigma\os)}\right.\right.\nonumber\\
&&\left.-\frac{3}{n}\left(\frac{V'(\sigma\os)}{V(\sigma\os)}\right)^2
-\frac{3X'(\sigma\os)}{\sigma\os}+\frac{3X(\sigma\os)}{\sigma\os^2}\right\}\nonumber\\
&&\left.\frac{V''(\sigma\os)}{V'(\sigma\os)}-(1-X(\sigma\os))\frac{V''(\sigma_k)}{V'(\sigma\os)}\right]\ .
\end{eqnarray}


\end{document}